\pdfoutput=1 
\documentclass[prl,aps,floatfix,amsmath,amssymb,showpacs,twocolumn]{revtex4}

\usepackage[pdftex]{graphicx}
\usepackage{color}
\usepackage{times}
\usepackage{graphicx}
\usepackage{amssymb}
\usepackage{bm} 
\usepackage{amsmath,amsfonts,latexsym}
\usepackage{hyperref}              

\newcommand*{\br}{\mathbf{r}}

 \newcommand{\braket}[1]{\ensuremath{\left\langle{#1}\right\rangle}}

\newcommand{\tc}[1]{\textcolor{black}{#1}}

\newcommand*{\bk}{\mathbf{k}}

\newcommand*{\phop}{\phi^{\vphantom{\dagger}}}
\newcommand*{\phdop}{\phi^\dagger}

\newcommand{\vect}[1]{\ensuremath{\mathbf{#1}}}

\begin{document}

\title{The phase diagram of 2D polar condensates in a magnetic field}
\date{\today}
\author{A. J. A. James and A. Lamacraft}
\affiliation{Department of Physics, University of Virginia,
  Charlottesville, VA 22904-4717, USA}
\begin{abstract}
Spin one condensates in the polar (antiferromagnetic) phase  in two dimensions are shown to undergo a transition of the Ising type, in addition to the expected Kosterlitz--Thouless (KT) transition of half vortices, due to the quadratic Zeeman effect. We establish the phase diagram in terms of temperature and the strength of the Zeeman effect using Monte Carlo simulations. When the Zeeman effect is sufficiently strong the Ising and KT transitions \tc{meet}. For very strong Zeeman field the remaining transition is of the familiar integer KT type.
\end{abstract}
\maketitle

Ultracold atomic gases represent a new frontier in quantum magnetism, where optical traps allow for the possibility of spontaneous ordering of the atoms' spin. In particular, systems of spinful bosons allow for the possibility of studying the interplay of Bose condensation (or superfluidity) and magnetic ordering in \emph{spinor condensates}. In this Letter we discuss one of the simplest such systems: the polar (or antiferromagnetic) spin-1 condensate in a magnetic field, realized in a gas of $^{23}$Na atoms \cite{Stenger:1998}. We show that there are two types of defects: vortices and domain walls, or strings. Some years ago a number of authors \cite{korshunov:1985,lee1985,carpenter1989phase} studied the interesting phase diagram that results from the competition between vortex interactions and string tension in a simple statistical mechanical model. Our goal here is to show that the same physics can arise in a atomic gas from quite different microscopic origins.


Consider a dilute gas of spin-1 bosons, described by a spinor $\bm{\phi}(\br)$.
The character of the magnetic order that develops at low temperature depends upon the interatomic interactions, described by two kinds of quartic terms~\cite{ho1998,ohmi1998}
\begin{align}\label{spin1Hint}
H_{\mathrm{int}}=\int d\br \left[\frac{c_0}{2}\left(\bm{\phi}^\dagger\bm{\phi}\right)^2+\frac{c_2}{2}\left(\bm{\phi}^\dagger\vect{S}\bm{\phi}\right)^2\right]
\end{align}
corresponding to a density-density and spin-spin interaction respectively.
Here $\vect{S}$ are the spin-1 angular momentum matrices. It is convenient to work in terms of Cartesian components, where $\left(S_i\right)_{jk}=-i\epsilon_{ijk}$. Writing $\bm{\phi}=\vect{a}+i\vect{b}$, where $\vect{a}$ and $\vect{b}$ are real vectors, the second term in Eq.~\eqref{spin1Hint} becomes $2c_2\left(\vect{a}\times\vect{b}\right)^2$. We see that for $c_2<0$ the energy is minimized at fixed density for $\vect{a}$, $\vect{b}$ perpendicular and equal, while for the case $c_2>0$ that will be our primary concern, $\vect{a}$ and $\vect{b}$ are parallel.
Conventionally these two possibilities are called the \emph{ferromagnetic} and \emph{polar} states, respectively. 

\begin{figure}
\includegraphics[width=0.45\textwidth]{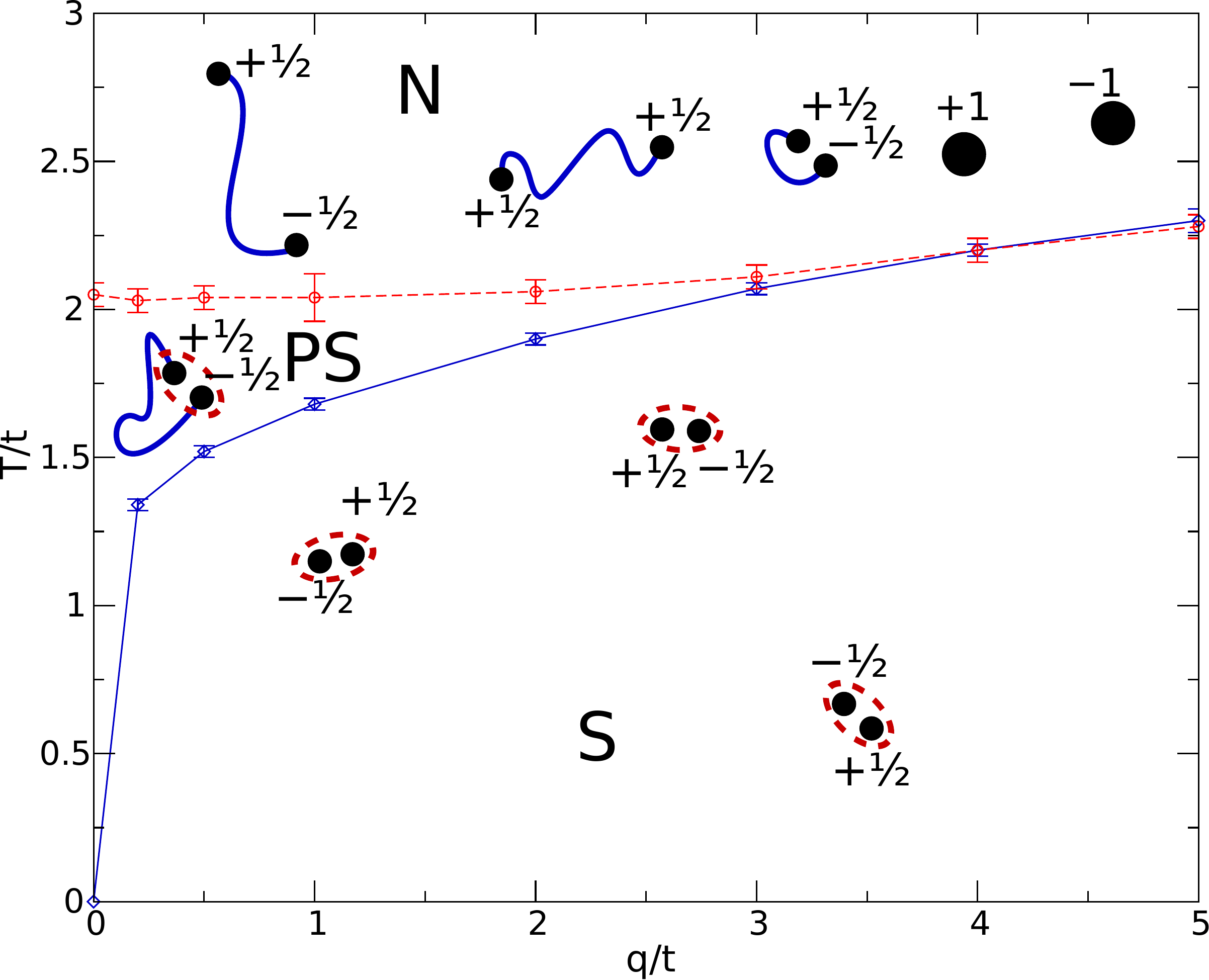}
\caption{(Color online) The phase diagram for a two dimensional polar condensate with quadratic Zeeman effect. The dashed red line and solid blue line mark the KT transition and Ising transition respectively. In the normal phase (N) half vortices of equal or opposite charge are joined by domain walls (blue lines). In the superfluid phase (S) vortices of opposite charge are bound (red ellipse) and there are no domain walls. In the pair superfluid phase (PS), \tc{$q/t \lesssim 4$,} vortices are bound but domain walls remain. \label{figphase}}
\end{figure}
In the mean-field description sketched above, a non-zero expectation value $\langle\bm{\phi}\rangle$, or equivalently off-diagonal long-range order in the density matrix $\rho_{ab}(\br,\br')\equiv\langle\phdop_a(\br)\phop_b(\br')\rangle$, simultaneously describes Bose condensation and the breaking of rotational symmetry. It is natural to ask whether these two phenomena necessarily go hand-in-hand, and if not, which occurs first as the temperature is lowered. We will show that in the two dimensional polar system ($c_{2}>0$, appropriate to $^{23}$Na), \tc{such an intermediate phase can arise}, possessing \emph{quasi}-long-range order in the singlet pair amplitude $\bm{\phi}\cdot\bm{\phi}$ -- a pair superfluid (PS) -- while the spin remains disordered. Spin ordering occurs in a \emph{second} transition at a lower temperature. This is possible \tc{when there is a small} anisotropy originating from the quadratic Zeeman effect, permitting an Ising transition where none would be allowed at zero field by the Mermin--Wagner theorem. \tc{At larger anisotropies the PS phase vanishes}. The resulting phase diagram, shown in Fig. \ref{figphase} for our Monte Carlo simulations of a particular model to be described shortly, is our principal  finding. The prospects for observing the transitions and the intermediate phase in an atomic gas will be discussed in the conclusion.

Let us begin by discussing the phases in Fig. \ref{figphase} in qualitative terms. The superfluid transition of scalar bosons in two dimensions is of the KT type, mediated by the binding of vortices, suggesting that we consider the analogous defects of a polar condensate. In the polar state we may write $\bm{\phi}= \vect{n}e^{i\theta}$, where $\vect{n}$ is a real unit vector and $\theta$ is a phase variable, (and we have we set the density equal to unity). In this representation taking $\vect{n} \to -\vect{n}$ and $\theta \to \theta+\pi$ maps the spinor to itself. Thus the elementary vortex has circulation $\frac{h}{2m}$, or one half of the usual quantum of circulation, and coincides with a `disclination' in the vector $\mathbf{n}$. 

The character of these point defects is dramatically altered by the inclusion of the Zeeman energy, which has the form
\begin{equation}
	\label{Zeeman}
	H_{\text{Z}}=\int d\br\, \bm{\phi}^{\dagger}\left[p S_{z}+q S_{z}^{2}\right]\bm{\phi}.
\end{equation}
Here $p$ and $q$ describe the linear and quadratic effects ($q>0$). We will be concerned with a system of zero total $S_{z}$, so that the linear term has no effect: the case of non-zero $S_{z}$ will be discussed briefly at the end. In the $(\mathbf{n},\theta)$ representation the quadratic effect contributes an energy per particle of $q (1-n_{z}^{2})$, amounting to an \emph{easy-axis} anisotropy for the $\mathbf{n}$ variable. 

At this point, it is convenient to introduce a simple lattice model \tc{which will be useful for our numerical simulations:}
\begin{align}\label{lattice_sigma_model}
H&=H_{\text{t}}+H_{\text{Z}}, \qquad H_{\text{Z}}=-q\sum_{i}n_{z,i}^{2}\nonumber\\
H_{\text{t}}&=-t\sum_{\langle ij\rangle}\left[\bm{\phi}^\dagger_{i}\cdot\bm{\phi}_j+\mathrm{c.c}\right]=-2t\sum_{\langle ij\rangle}\vect{n}_i\cdot\vect{n}_j\cos\left(\theta_i-\theta_j\right)
\end{align}
with hopping parameter $t$. \tc{The corresponding continuum model for the spinor} takes the form
\begin{align}\label{cont_sigma}
H\to \int d^d\br \left[t\left(\nabla\vect{n}\right)^2+t\left(\nabla\theta\right)^2-qn_{z}^{2}\right]
\end{align}
(we take the lattice spacing equal to unity). Notice that the superfluid and magnetic degrees of freedom appear to decouple in this expression. The only coupling is global, in that half-vortex / disclinations are allowed. Thus when these defects are absent (or bound) the $\mathbf{n}$ degrees of freedom are described by the familiar Heisenberg model with anisotropy.

\begin{figure}
	\centering
		\includegraphics[width=0.35\textwidth]{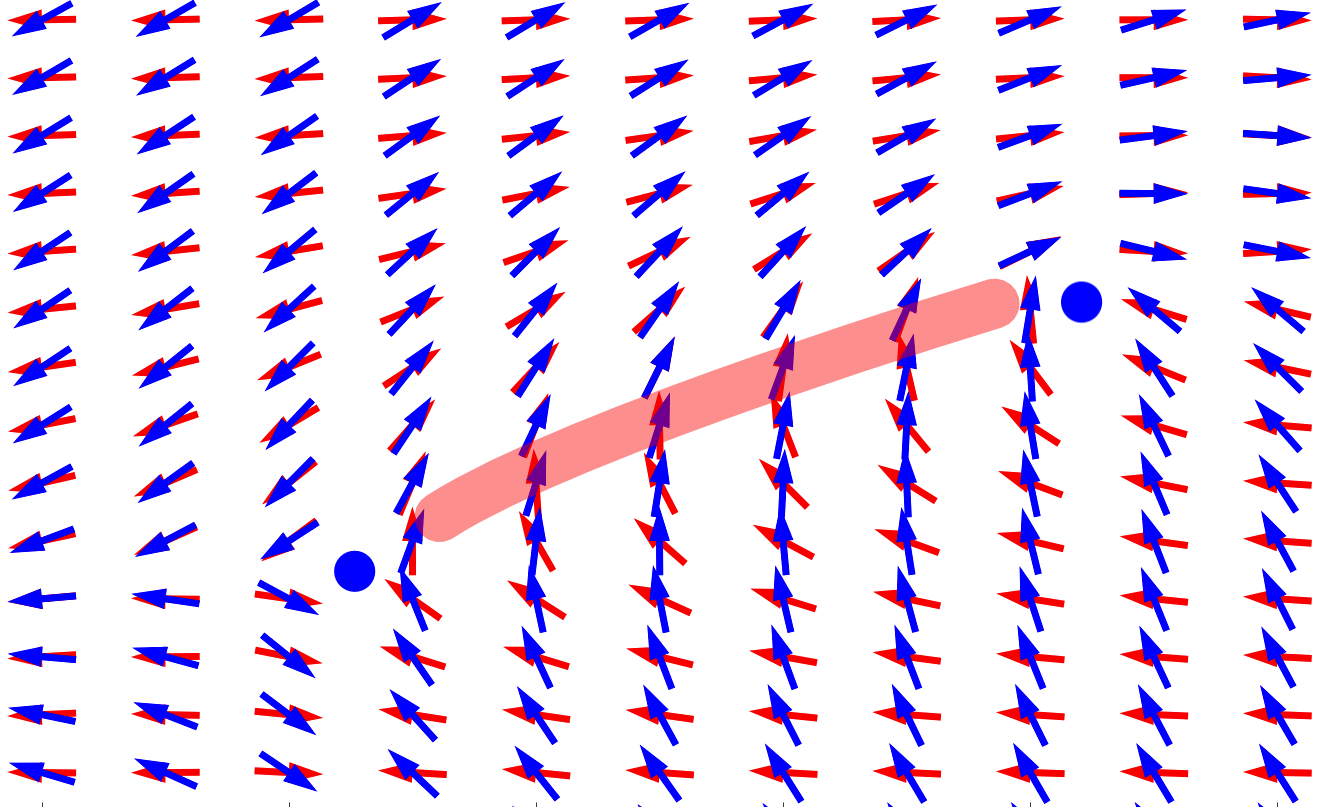}
	\caption{(Color online) A half vortex pair connected by a domain wall (shaded region) in the $\vect{n}$ vector (shown red) arising from easy axis anisotropy (the easy axis is horizontal). The blue arrows denote the configuration of the phase $\theta$.}
	\label{fig:DomainPair}
\end{figure}

In Eq.~\eqref{cont_sigma}, $q$ appears as a `mass' for deviations from the easy axis, meaning that such deviations are confined to `domain walls' of thickness $\propto q^{-1/2}$ that have an energy per unit length -- or tension -- $\propto q^{1/2}$. Furthermore, these domain walls can \emph{terminate} on the half-vortices discussed above, see Fig. \ref{fig:DomainPair}.

Now we are in a position to understand the structure of the phase diagram. At $q=0$, the only finite temperature transition is a KT transition at $T=T_{KT}\approx t \pi/8$ due to binding of the half-vortices, which have lower energy than integer vortices \cite{mukerjee2006topological,zhou2001spin}. The distinctive features of this transition will be discussed later. After the half-vortices become bound, the description of the $\mathbf{n}$ degrees of freedom coincides with that of the ordinary Heisenberg model, and no order appears at finite temperature. For $q$ nonzero but small the domain walls connecting half-vortices have too little tension to affect the KT transition. Once the half-vortices have become bound, the domain walls are closed but otherwise fluctuating, and can disappear in an Ising transition at a lower temperature. An alternative way to think about this second transition is in terms of the Heisenberg model, which has an Ising transition in the presence of an easy-axis anisotropy. Finally, at large $q$, the $\mathbf{n}$ vector is pinned to the $\pm \hat{\mathbf{z}}$ direction and the model Eq.~\eqref{lattice_sigma_model} coincides with the usual XY model after shifting $\theta_{i}\to \theta_{i}+\pi n_{z,i}/2$. In this regime there is a single KT transition of integer vortices. In terms of the original spin states of the boson, only the $S_{z}=0$ state is occupied, so that the behavior of a scalar condensate is recovered. An earlier investigation considered only $q=0$ and large $q$, so that the interesting interpolation between these two limits went unnoticed \cite{mukerjee2006topological}. We note parenthetically that the case $q<0$ -- harder to realize experimentally -- was also studied recently \cite{Podolsky:2009}.

The region occupied by the intermediate PS phase in the model Eq.~\eqref{lattice_sigma_model} is very small: its presence could not be unambiguously determined by Monte Carlo simulations on systems of up to $40 \times 40$ sites.
This is likely due to the Ising transition line being very steep near $q=0$. Standard arguments for the scaling of the `mass' (correlation length) with temperature \cite{chaikin2000principles} show that $T \sim -1/\log(q)$ for small $q$.
We therefore study a generalized model with a `pair hopping' term 
\begin{align}\label{pair_hopping}
H_{\text{u}}=-\frac{u}{2}\sum_{\braket{ij}}\left(\bm{\phi}^{\dagger}_{i}\cdot\bm{\phi}^{\dagger}_{i}\right)\left(\bm{\phi}^{\vphantom{\dagger}}_{j}\cdot\bm{\phi}^{\vphantom{\dagger}}_{j}\right)+\text{c.c.}
\end{align}
Including $H_\text{u}$ in Eq.~\eqref{cont_sigma}, we see that a finite $u$ only stiffens the phase variable, changing the coefficient of the $\theta$ term from $t$ to $t+2u$. The half-vortex KT transition then occurs at the higher temperature $T \approx (t+2u)\pi/8$, increasing the size of the PS phase. In the following we take $u=2t$ as then the \tc{PS phase} is clearly visible even for moderate system sizes. Experimentally a similar result could be achieved by increasing $c_2$ until 2-body singlet bound states form at $q=0$. Then $\mathbf{n}$ is disordered even at $T=0$ (enlarging the PS phase) until $q$ is large enough to cause a \emph{quantum} phase transition into the S phase. Photoassociation data suggest that the required condition, a divergent singlet scattering length, has been achieved already via optical Feshbach resonance (Fig. 7 of Ref. \cite{opticalfeshbach}).

For $q\to\infty $ the model is equivalent to the Hamiltonian of the generalised XY model (see e.g.\cite{korshunov:1985,lee1985,carpenter1989phase})
\begin{align}
 H_{\Delta}&=- \sum_{\braket{ij}} \Big( \Delta\cos(\theta_i-\theta_j) +(1-\Delta)\cos(2\theta_i - 2\theta_j)\Big), \label{eqhcc}
\end{align}
where conventionally $\Delta \in [0,1]$. This model also exhibits a PS phase. Our choice of parameters corresponds to the case $\Delta=0.5$ so that, according to the phase diagram in Ref.~\cite{carpenter1989phase} the large $q$ limit of our model with $u=2t$ \emph{still} has a single transition (hence $u$ itself does not produce a PS phase for $q\to\infty$). As $q$ is reduced this transition should split in two.

We study the phase diagram of the model $H=H_{\text{t}}+H_{\text{u}}+H_{\text{Z}}$ via Monte Carlo simulations (using tools from the ALPS libraries \cite{albuquerque2007alps}) on square systems of $L \times L$ sites with periodic boundary conditions. As there are three continuous parameters per site a large number of sweeps of the lattice  are required to equilibrate and collect reliable data, even for small system sizes ($> 10^6$ for $L=8$). We performed simulations for $L=8,16,24,32$ with some extra data collected for $L=40,48$ in special cases.
To detect two separate transitions, we consider the specific heat capacity in addition to the Binder cumulants for both the spinor and the phase. 

\begin{figure}
\includegraphics[width=0.45\textwidth]{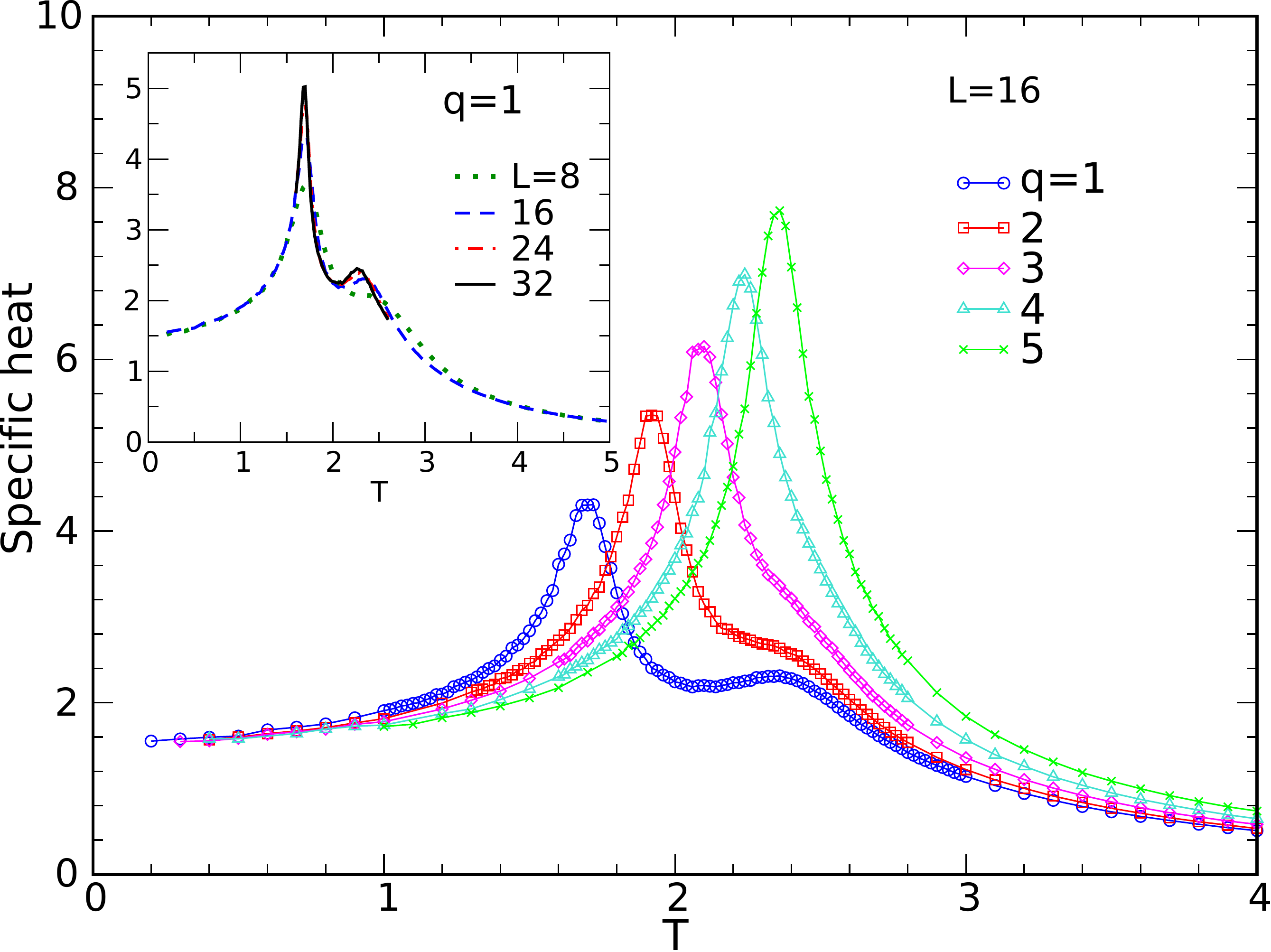}
\caption{(Color online) The specific heat, $C$, for $L=16$, showing the merging of the peaks associated with the two transitions as $q$ increases. Inset: $C$ for different system sizes at $q=1$. \label{figheats}}
\end{figure}

A phase transition of the Ising type at $T=T_c$ should present itself as a sharp peak in the specific heat, $C$, for finite size simulations. KT transitions are also accompanied by a peak in $C$, above $T_{KT}$, associated with the increase in entropy when vortices unbind.
Fig. \ref{figheats} shows that as $q$ approaches a critical value, $q \sim 4$, the sharper lower temperature Ising peak and the broader higher temperature KT peak merge.

\begin{figure}
\includegraphics[width=0.45\textwidth]{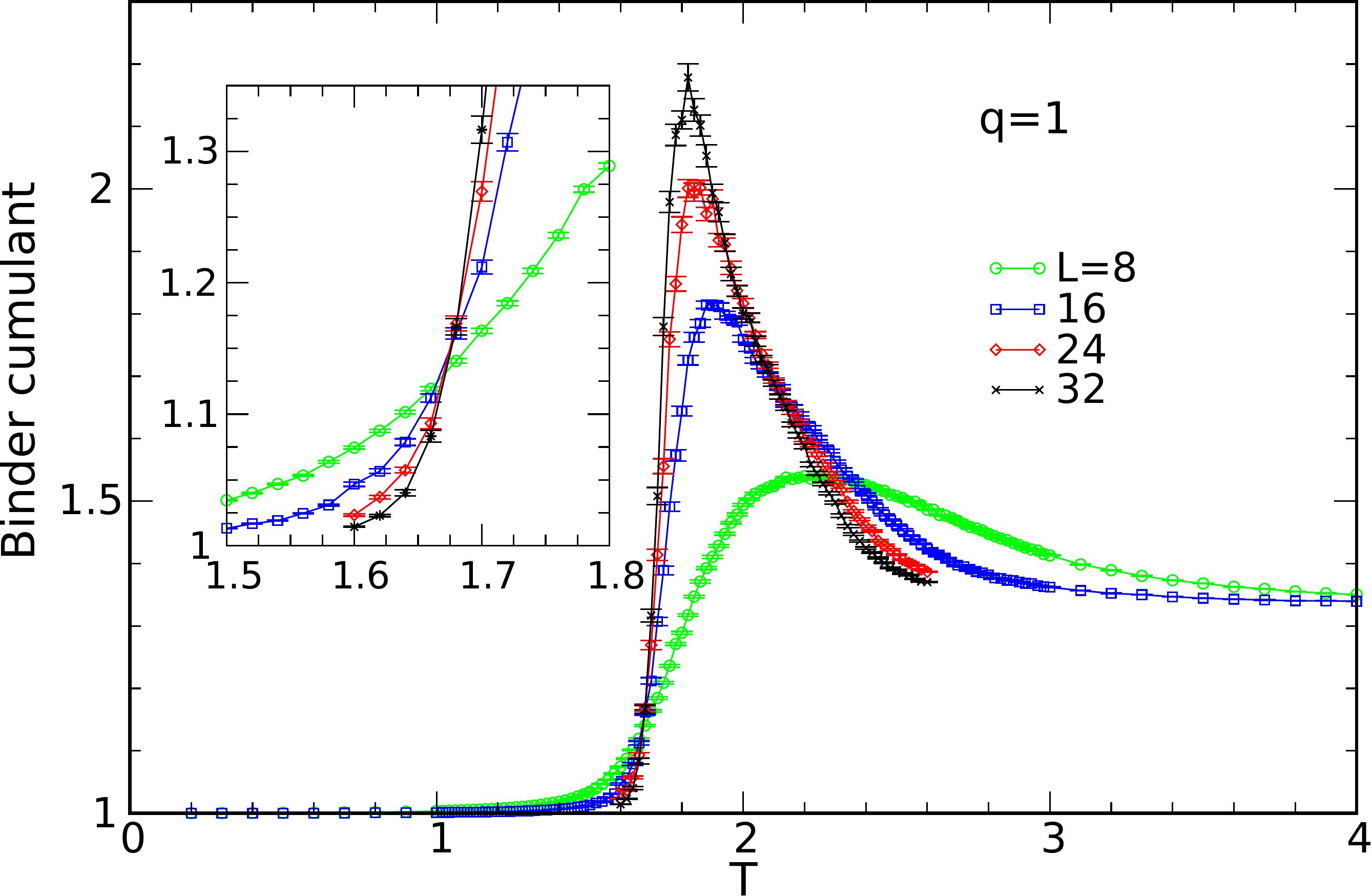}
\caption{(Color online)  Intersection of the Binder Cumulants, $B[\phi]$ for different $L$ at $q=1$. Inset: enlarged region of intersection. \label{figBinder}}
\end{figure}

Better quantitative information is given by the Binder cumulants \tc{\cite{binder1981finite}}.
The Binder cumulant for the spinor is
\begin{align}
B[\phi]=\frac{\braket{(\vect{\Phi}^\dagger\cdot\vect{\Phi})^2}}{\braket{(\vect{\Phi}^\dagger\cdot\vect{\Phi})}^2} \label{eqbinder}
\end{align}
where $\vect{\Phi}=\sum_i \vect{\phi}_i /N$. An example plot is given in Fig. \ref{figBinder}.
We also calculate the cumulant for the $z$ component:
\begin{align}
B[\phi_z]=\frac{\braket{(\Phi_z^\dagger\Phi_z)^2}}{\braket{(\Phi_z^\dagger\Phi_z)}^2}.\label{eqbinderz}
\end{align}
These cumulants are sensitive to order and Ising-like order in $\phi$, respectively. 
On the other hand they are not sensitive to order (or quasi long-range order) in the phase alone. Instead, to look for order in $\exp(2 i\theta)$, we use the cumulant
\begin{align}
B[2\theta]=\frac{\sum_{ijkl}\braket{ \exp{[2i(\theta_i-\theta_j+\theta_k-\theta_l)]}}}{(\sum_{ij}\braket{ \exp{[2i(\theta_i-\theta_j)]}})^2}.
\end{align}

In the vicinity of a conventional, continuous transition at $T_c$, and where finite size scaling holds, the Binder cumulant for a suitable variable can be written in the form
\begin{align}
B_L=\tilde{B}(\tilde{T}L^{\frac{1}{\nu}}) \label{eqBL}
\end{align}
where $\tilde{B}$ is a universal scaling function, and $\tilde{T}=(T-T_c)/T_c$. From this we conclude that Binder cumulants for different $L$ cross at $\tilde{T}=0$, providing an accurate method for determining $T_c$.
For KT transitions, eq. (\ref{eqBL}) does not hold. However the crossings for different $L$ still occur in a suitably narrow range \cite{loison1999binder} (see Fig. \ref{figBinderd}), allowing us to estimate $T_{KT}$.
\begin{figure}
\includegraphics[width=0.42\textwidth]{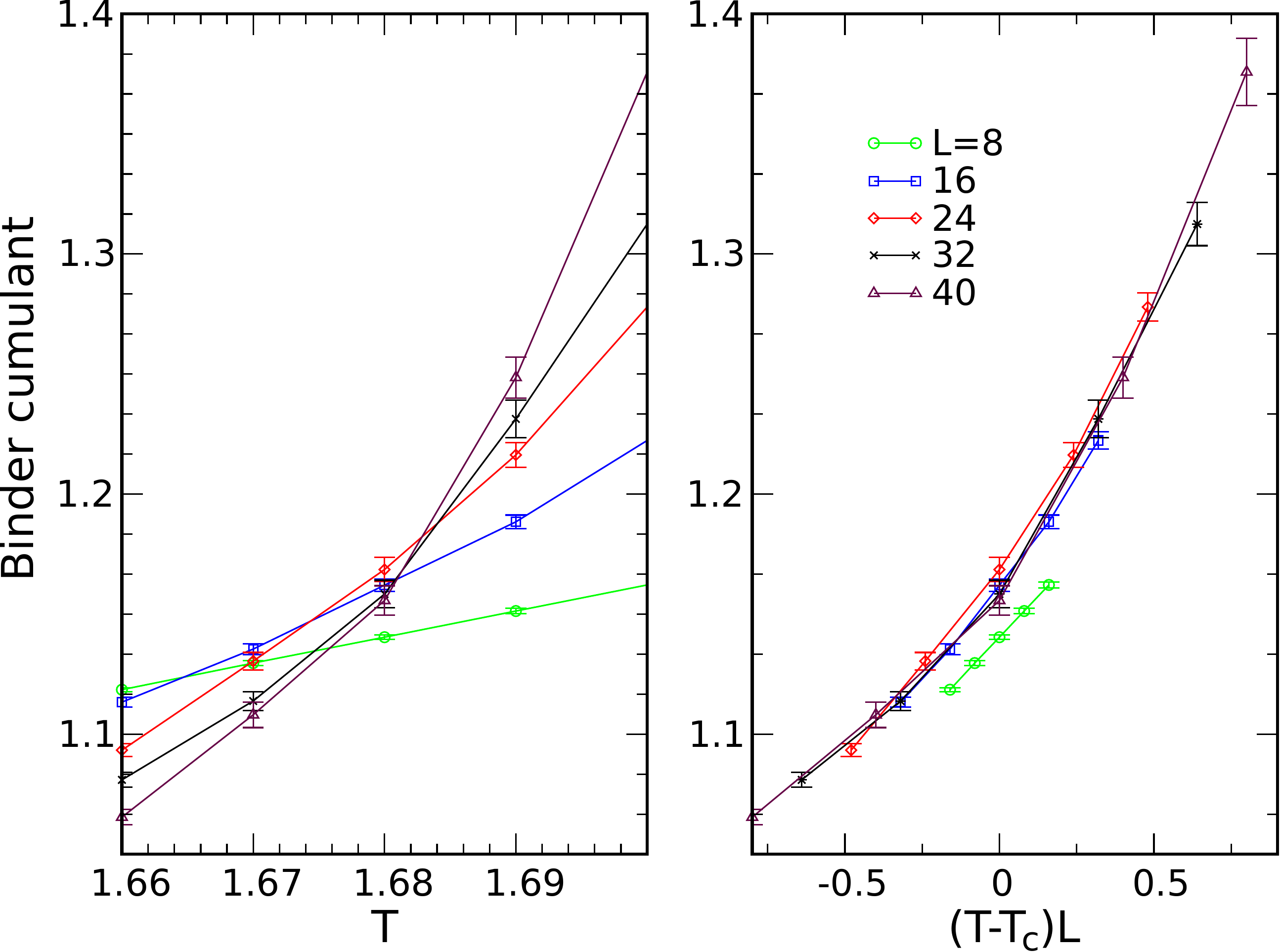}
\caption{(Color online)  Left: Intersection of the Binder Cumulants $B[\phi]$ for different $L$ at $q=1$. Right: The same data but \tc{plotted against $(T-T_c)L$}. Note that the data for $L=8$ is not within the range of validity for finite size scaling.\label{figscaled}}
\end{figure}
\begin{figure}
\includegraphics[width=0.45\textwidth]{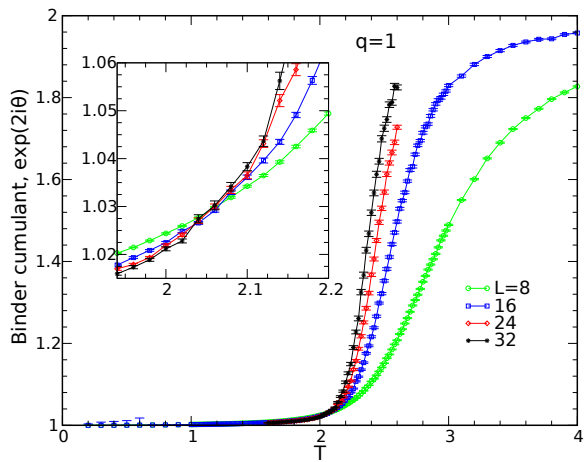}
\caption{(Color online)  Intersection of the Binder Cumulants $B[2\theta]$ for different $L$ at $q=1$. Inset: enlarged region of intersection. \label{figBinderd}}
\end{figure}
Fig. \ref{figphase} shows $T_c$ and $T_{KT}$ as found using the above cumulants. The values of $T_c$ provided by $B[\phi]$ and $B[\phi_z]$ agree within error over the full range of $q$ investigated. As a further check on the nature of the PS to S transition we extract the critical exponent $\nu$ by calculating $dB_L/dT$ at $T_c$. We find that $\nu$ is at least consistent with the Ising value $\nu=1$, within error. An example of the resulting data collapse is shown in Fig. \ref{figscaled}. 

\begin{figure}
\includegraphics[width=0.45\textwidth]{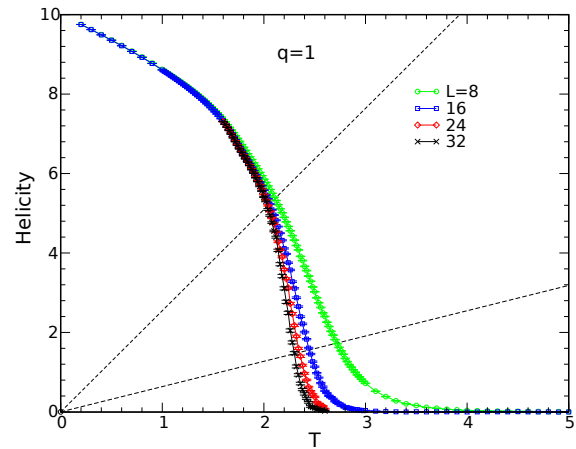}
\caption{(Color online)  Helicity for $q=1$. The upper and lower dashed lines are $\frac{8T}{\pi}$ and $\frac{2T}{\pi}$ respectively. The jump is clearly commensurate with $\Delta\Upsilon=\frac{8T}{\pi}$. \label{fighelicity}}
\end{figure}

Following Ref. \cite{mukerjee2006topological} we also examine the helicity modulus (also known as spin stiffness or superfluid density), $\Upsilon$, defined as the change in free energy due to a twist in boundary conditions along some direction, $\hat{\vect{x}}$. 
A KT phase transition at $T_{KT}$ is reflected in the helicity in the form of a jump proportional to $T_{KT}$ (see e.g. \cite{chaikin2000principles}). When the transition is driven by half-integer vortices this jump will be four times larger than for the integer case \cite{korshunov:1985a}:
\begin{align}
\Delta\Upsilon_{\frac{1}{2}KT}=4\Delta\Upsilon_{KT}=\frac{8T_{KT}}{\pi}.
\end{align}
The observed position of the helicity jump as a function of $q$ and $T$ confirms that the transition temperature provided by $B[2\theta]$ represents $T_{KT}$ with reasonable accuracy. As in Ref.~\cite{mukerjee2006topological} we find that the N to PS transition is due to the presence of half vortices (Fig. \ref{fighelicity}). In fact, setting $u=2t$ means that the transition is facilitated by half vortices even for $q \to \infty$. This is consistent with eq. (\ref{eqhcc}) for $\Delta=0.5$.  In that case half-vortices still exist, 
\tc{though they occur in the phase alone and the line defects that join them have energy independent of $q$} \cite{carpenter1989phase}. As $u$ approaches zero the integer KT transition at large $q$ should be recovered.

We now return to the case of finite $S_z$. As in the case of an antiferromagnet with an easy axis anisotropy, increasing $S_{z}$ leads to a spin-flop transition between a state with $\vect{n}$ aligned in the $z$-direction to one where it lies in the $x-y$ plane. Such a transition is described by a bicritical point of the Heisenberg type, which in $d=2$ must occur at $T=0$ \cite{fisher1974spinflop,zhou2006hidden}.
At finite $T$, the low $S_{z}$ Ising and high $S_{z}$, $xy$ ordered, phases are separated by a normal region.
The high $S_{z}$ phase resembles the $q<0$ case discussed in Ref.~\cite{Podolsky:2009}.

In conclusion we have argued that polar condensates undergo separate KT and Ising transitions when subjected to the quadratic Zeeman effect. We have supported this finding with Monte Carlo simulations. Aside from the thermodynamic signatures discussed here, the PS phase should be visible in the correlation function of occupancies of different momentum states, as measured by the noise correlations in an image of the expanded gas: $\langle\delta n(\bk_{1})\delta n(\bk_{2})\rangle\propto |\bk_{1}+\bk_{2}|^{4\eta-2}$ with $\eta=T/(2\pi \Upsilon)$ \cite{altman:2004}. 

The authors acknowledge support of the NSF under grant DMR-0846788 and wish to think Chris Dawson for his helpful comments regarding the simulations.


\end{document}